\documentclass[aps,twocolumn]{revtex4}
\usepackage{epsfig}
\newcommand{\be}{\begin{equation}}
\newcommand{\ee}{\end{equation}}
\newcommand{\ba}{\begin{array}}
\newcommand{\ea}{\end{array}}
\newcommand{\tq}{\tilde{q}}

\begin{document}
\title{Growth Algorithms for Lattice Heteropolymers at Low Temperatures}
\author{Hsiao-Ping Hsu, Vishal Mehra, Walter Nadler, and Peter Grassberger}
\affiliation{John-von-Neumann Institute for Computing, Forschungszentrum
J\"ulich, D-52425 J\"ulich, Germany}

\date{\today}
\begin{abstract}
Two improved versions of the pruned-enriched-Rosenbluth method (PERM) are
proposed and tested on simple models of lattice heteropolymers. Both are
found to outperform not only the previous version of PERM, but also all
other stochastic algorithms which have been employed on this problem, except
for the core directed chain growth method (CG) of Beutler \& Dill. In nearly
all test cases they are faster in finding low-energy states, and in many
cases they found new lowest energy states missed in previous papers. The
CG method is superior to our method in some cases, but less efficient in
others. On the other hand, the CG method uses heavily heuristics based on
presumptions about the hydrophobic core and does not give thermodynamic
properties, while the present method is a fully blind general purpose
algorithm giving correct Boltzmann-Gibbs weights, and can be applied in
principle to any stochastic sampling problem.
\end{abstract}
\maketitle

\section{Introduction}

Lattice polymers have been studied intensively to understand phenomena like
the globule-coil transition of polymers, protein folding, etc.
Protein folding (or, more precisely, protein fold prediction), one of the
central problems of computational biology, refers to the determination
of the ground state of protein molecules -- which grosso modo is also its
native state -- from their amino acid sequence. Due to rapid advances in
DNA analysis the number of known
sequences has increased enormously, but progress in understanding their
3-dimensional structure and their functions has lagged behind owing
to the difficulty of solving the folding problem.

Simplifying the description of a protein by replacing each amino acid by a
simple point particle on a site of a regular lattice implies of course a
great reduction of complexity, and one might wonder how much one can learn
by this for real proteins. But even if this simplification is too strong,
searching for lowest energy states of such models represents a paradigmatic
example of combinatorial optimization. This will indeed be our main
motivation: Finding algorithms that explore efficiently the low-energy
states of a complicated energy landscape with many local minima.
In addition to finding the ground state we want these algorithms also
to sample excited states correctly, so that they provide a complete
thermodynamic description -- though we shall restrict ourselves in this
paper to presenting results on ground states only.

A popular model used in these studies is the so-called HP
model \cite{Dill85,Lau89} where only two types of monomers,
H (hydrophobic) and P (polar) ones, are considered. Hydrophobic monomers
tend to avoid water which they can only by mutually attracting themselves.
The polymer is modeled as a self-avoiding chain on a regular (square or
simple cubic) lattice with repulsive or attractive interactions between
neighboring non-bonded monomers. Although also other interaction parameters
have been used in the literature, almost all examples treated in this paper
use energies $\epsilon_{HH} = -1, \;\epsilon_{HP}=\epsilon_{PP}=0$. The
only other model studied here has also two types of monomers, for simplicity
also called H and P (although they have identical hydrophobicities),
but with $\epsilon_{HH} = \epsilon_{PP}=-1,\; \epsilon_{HP}=0$ \cite{tp92}.
Chain lengths considered in the literature typically are between $N=30$ and
$N=100$. Shorter chains do not present any problem, longer ones are too
difficult.

A wide variety of computational strategies have been employed to simulate and
analyze these models, including conventional (Metropolis) Monte Carlo
schemes with various types of moves \cite{yue95,rrp97,deutsch97}, chain growth
algorithms without \cite{tp92} and with re-sampling \cite{g98,bast98} (see
also \cite{bd96}),
genetic algorithms \cite{um93,kd01}, parallel tempering \cite{irbaeck}, and
generalizations thereof \cite{chikenji,chikenji2}, an `evolutionary Monte
Carlo' algorithm \cite{lw01}, and others \cite{tt96}. In addition, Yue and
Dill \cite{yd93,yd95} also devised an exact branch-and-bound algorithm
specific for HP sequences on cubic lattices which gives all low energy states
by exact enumeration, and typically works for $N{<\atop\sim} 70 - 80$. If
the chain is too long, it does not give wrong output but no output at all.

It is the purpose of the present letter to present two new variants of the
Pruned-Enriched Rosenbluth Method (PERM) \cite{g97} and to apply them to
lattice proteins. PERM is a biased chain growth algorithm with re-sampling
(``population control") with depth-first implementation. It has a certain
resemblance to genetic algorithms, except that the latter are usually
implemented breadth-first and do not allow to obtain correct Gibbs-Boltzmann
statistics.

The original version of PERM was used for lattice protein folding in
\cite{g98,bast98} and did extremely well. With one exception, it could
find all known lowest energy configurations for all sequences tested in
\cite{g98,bast98}, and found a number of new lowest energy states. The one
case where it could not find the ground state in an unbiased and blind
search was a 64-mer designed in \cite{um93} (see Fig.~1), but this is not
surprising: Any chain growth algorithm should have problems in finding this
configuration, since it has to grow a long arc which at first seems very
unnatural and which is stabilized only much later. Indeed, at that time no
other Monte Carlo method had been able to find this state either. But a very
efficient algorithm, the {\it Core-directed Growth method} (CG) \cite{bd96}
was overlooked in \cite{g98,bast98}. Thus PERM was not tested on the most
difficult example known at that time, a 88-mer forming a $\beta/\alpha$-barrel
whose ground state energy was known exactly. In the meantime, also two
other improved Monte Carlo algorithms were published \cite{irbaeck,chikenji}.
All this motivated us to take up the problem again.

\begin{figure}
  \begin{center}
   \psfig{file=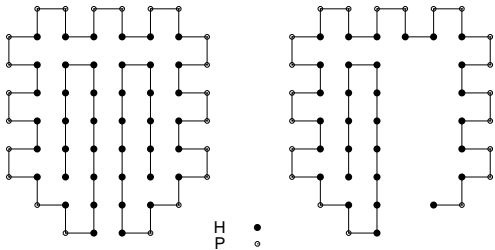,width=5.0cm,
angle=270}
   \vskip 0.3 true cm
\caption{Left side: ground state configuration of a $N=64$ chain in 2D from
\cite{um93}.
   Other states with the same energy differ in the detailed folding of the
   tails in the interior, but have identical outer shapes. Right side: when
   about 3/4 of the chain is grown, one has to pass through a very
   unstable configuration which is stabilized only later, when the core is
   finished.}
\label{2d-64}
\end{center}
\end{figure}

\section{The Algorithm}

PERM is built on the old idea of Rosenbluth and Rosenbluth (RR) \cite{rr55}
to use a biased growth algorithm for polymers, where the bias is corrected
by means of giving a weight to each sample configuration. While the chain
grows by adding monomers, this weight (which also includes the Boltzmann
weight if the system is thermal) will fluctuate. PERM suppresses
these fluctuations by pruning configurations with too low weight, and
by ``enriching" the sample with copies of high-weight configurations
\cite{g97}. These copies are made while the chain is growing, and continue
to grow independently of each other. PERM has been applied successfully to a
wide class of problems, including e.g. the $\Theta$ transition in homopolymers
\cite{g97}, trapping of random walkers on absorbing lattices \cite{mg02},
and stretching collapsed polymers in a poor solvent \cite{gh02}. It can
be viewed as a special realization of a ``go with the winners" strategy
\cite{aldous} which indeed dates back to the beginning of the Monte Carlo
simulation era, when it was called ``Russian roulette and splitting"
\cite{kahn}. Among statisticians, this approach is also known as
sequential importance sampling (SIS) with re-sampling \cite{liu}.

Pruning and enrichment were done in \cite{g97,g98,bast98} by choosing
thresholds $W_n^<$ and $W_n^>$ depending on the estimate of the partition sums
of $n$-monomer chains. These thresholds are continuously updated as the
simulation progresses.
If the current weight $W_n$ of an $n$-monomer chain is less than $W_n^<$,
a random number $r$ is chosen uniformly in $[0,1]$. If $r<1/2$, the chain
is discarded, otherwise it is kept and its weight is doubled. Thus low-weight
chains are pruned with probability 1/2. Many alternatives to this simple
choice are discussed in \cite{liu}, but we found that more sophisticated
strategies had little influence on the efficiency, and thus we kept the
above in the present work. The determination of $W_n^<$ and $W_n^>$ will
be discussed later. In principle we could use the same as in
\cite{g98,bast98}, but we simplified it since the new variants are more
robust, and some of the tricks employed in \cite{g98,bast98} are not needed.

On the contrary, we found that different strategies in biasing and, most
of all, in enrichment had a big effect, and it is here the
present variants differ from those in \cite{g98,bast98}. There, high-weight
configurations were simply cloned (with the number of clones determined
from the ratio of the actual weight to $W_n^>$),
and the weight was uniformly shared between the clones.
For relatively high temperatures
this is very efficient \cite{g97}, since each clone has so many possibilities
to continue that different clones very quickly become independent from each
other.
This is no longer the case for very low temperatures. There we found that
clones
often evolved in the same direction, since one continuation has a much higher
Boltzmann weight than all others. Thus, cloning is no longer efficient in
creating configurational diversity, which was the main reason why it was
introduced.

The main modification made in the present paper is thus that we no longer
make {\it identical clones}. Rather, when we have a configuration with $n-1$
monomers, we first estimate a {\it predicted} weight $W_n^{\rm pred}$ for
the next step, and we count the number $k_{\rm free}$ of free sites where
the $n$-th monomer can be placed. If $W_n^{\rm pred}> W_n^>$ and $k_{\rm
free}>1$,
we choose $2 \le k\le k_{\rm free}$ {\it different} sites among the free
ones and continue with $k$ configurations which are {\it forced} to be
different. Thus we avoid the loss of diversity which limited the success of
old PERM. We tried several strategies for selecting $k$ which all gave
similar results. Typically, we used $k = \min \{k_{\rm free},
\lceil W_n^{\rm pred}/W_n^>\rceil\}$.

When selecting a $k$-tuple $A = \{\alpha_1,\ldots \alpha_k\}$ of mutually
different continuations $\alpha_j$ with probability $p_A$,
the corresponding weights $W_{n,\alpha_1}\ldots, W_{n,\alpha_k}$ are
(see Appendix)
\be
   W_{n,\alpha_j} = \frac{ W_{n-1} q_{\alpha_j} k_{\rm free}}
                     { k {k_{\rm free}\choose k} p_A}\;.
\label{nis0}
\ee
Here, the {\it importance}
\be
   q_{\alpha_j} = \exp(-\beta E_{n,\alpha_j})            \label{qn}
\ee
of choice $\alpha_j$ is the Boltzmann-Gibbs factor associated with the
energy $E_{n,\alpha_j}$ of the newly placed monomer in the potential created
by all previous monomers, and the terms in the denominator of Eq.~(\ref{nis0})
arise from correcting bias and normalization.

For the choice of continuations among the $k_{\rm free}$ candidates, we
used two different strategies:

(1) In the first, called nPERMss for ``new PERM with simple sampling", we
chose them randomly and uniformly,
with the only restriction that they are mutually different.
Accordingly, $W_n^{\rm pred} = W_{n-1} k_{\rm free}$ \cite{footnote1}, and
\be
   W_{n,\alpha} = W_{n-1} q_\alpha k_{\rm free} /k\;.
\ee
This has the advantage of simplicity,
but it might at first appear to be inefficient. A priori, we would
expect that some bias in favour of continuations with high Boltzmann weights
or against continuations which run into dead ends might be necessary for
efficiency.

(2) In the second, called nPERMis for ``new PERM with importance sampling",
we did just that. For each possible placement $\alpha\in [1,k_{\rm free}]$
of the $n$-th monomer we calculated its energy $E_{n,\alpha}$ and its number
$k_{\rm free}^{(\alpha)}$ of free neighbours, and used modified importances
defined by
\be
   \tq_\alpha = (k_{\rm free}^{(\alpha)} + 1/2) \exp(-\beta E_{n,\alpha})
\ee
to choose among them. The predicted weight is now $W_n^{\rm pred} = W_{n-1}
\sum_\alpha \tq_\alpha$. The replacement of $q_\alpha$ by $\tq_\alpha$ is
made since we anticipate that continuations with less free neighbours will
contribute less on the long run than continuations with more free neighbours.
This is similar to ``Markovian anticipation" \cite{cylinder} within
the framework of old PERM, where a bias different from the short-sighted
optimal importance sampling was found to be preferable.

The actual choice was made such that, for a given $k$ (remember that $k$
was already fixed by the ratio $W_n^{\rm pred}/W_n^>$), the variance of the
weights $W_n$ is minimal. For $k=1$ this is standard importance sampling,
$p_\alpha = \tq_\alpha/\sum_{\alpha'} \tq_{\alpha'}$, and the variance of
$W_n$ for fixed $W_{n-1}$ would be zero if we had not replaced $q_\alpha$
by $\tq_\alpha$: $W_{n,\alpha}=W_{n-1} q_\alpha/p_\alpha = W_{n-1} q_\alpha
/\tq_\alpha \sum_{\alpha'} \tq_{\alpha'}$.
For $k>1$, the probability to select a tuple $A = \{\alpha_1,
\ldots \alpha_k\}$ is found to be
\be
   p_A = {\sum_{\alpha \in A} \tq_\alpha \over \sum_{A'}\sum_{\alpha'\in A'}
         \tq_{\alpha'}}\;.                                     \label{nis}
\ee
The corresponding weights are determined according to Eq.~(\ref{nis0}).
The variance of the weight increase $W_{n,\alpha}/W_{n-1}$, summed over all
$k$ continuations within the tuple, would again be zero if $q_\alpha$ were
not replaced by $\tq_\alpha$.

nPERMis is more time consuming than nPERMss, but it should also be more
efficient. While Eq.~(\ref{nis}) with $\tq_{\alpha}$ replaced by $q_{\alpha}$
would be optimal if the chain growth were a Markov process, it is not
guaranteed to be so in the actual (non-Markovian) situation. We tried
some alternatives for $p_A$, but none gave a clear improvement.

A noteworthy feature of both nPERMss and nPERMis is that they cross over to
complete enumeration when $W_n^<$ and $W_n^>$ tend to zero. In this limit, all
possible branches are followed and none is pruned as long as its weight is
not strictly zero. In contrast to this, old PERM would have made exponentially
many copies of the same configuration. This suggests already that we can be
more lenient in choosing $W_n^<$ and $W_n^>$. For the first configuration
hitting
length $n$ we used $W_n^<=0$ and $W_n^>=\infty$, i.e. we neither pruned nor
branched. For the following configurations we used $W_n^>=CZ_n/Z_0 (c_n/c_0)^2$
and $W_n^<=0.2\,W_n^>$. Here, $c_n$ is the total number of configurations
of length $n$ already created during the run, $Z_n$ is the partition sum
estimated from these configurations, and $C$ is some positive number $\le 1$.
The following results were all obtained with $C=1$, though substantial
speed-ups (up to a factor 2) could be
obtained by choosing $C$ much smaller, typically as small as $10^{-15}$ to
$10^{-24}$. The latter is easily understandable: with such small $C$, the
algorithm performs essentially exact enumeration for short chains, giving
thus maximal diversity, and becomes stochastic only later when following
all possible configurations would become unfeasible. We do not quote the
optimal results since they are obtained only for narrow ranges of $C$ which
depend on the specific amino acid sequence, and finding them in each case
would require an extensive search.

Since both nPERMss and nPERMis turned out to be much more efficient and robust
than old PERM, we did not use special tricks employed in \cite{g98} like
growing
chains from the middle rather than one of the ends, or forbidding contacts
between polar monomers.

In the following, when we quote numbers of ground state
hits or CPU times between such hits, these are always {\it independent} hits.
In PERM we work at a fixed temperature (no annealing), and successive ``tours"
\cite{g97} are independent except for the thresholds $W_n^{<,>}$ which use
partially the same partition sum estimates. The actual numbers of (dependent)
hits are much larger.

For both versions, results are less sensitive to the precise choice of
temperature
than they were for old PERM. As a rule, optimal results were obtained at
somewhat lower temperatures, but in general all temperatures in the range
$0.25 < T < 0.35$ gave good results for ground state search.


\section{Results}

{\bf (a)} We first tested the ten 48-mers from \cite{yue95}. As with old PERM,
we could reach lowest energy states for all of them, but within much shorter 
CPU times. As seen from Table~\ref{table1}, nPERMis did slightly better than
nPERMss, and both were about one order of magnitude faster than old PERM. For 
all 10 chains we used the same temperature, $\exp(1/T) = 18$, although we 
could have optimized CPU times by using different temperatures for each 
chain. In the following we quote in general only results for \hbox{nPERMis}, 
but results for nPERMss were nearly as good.

The CPU times for nPERMis in Table~\ref{table1} are typically one order of
magnitude smaller than those in \cite{bd96}, except for sequence \#9 whose
lowest energy was not hit in \cite{bd96}. Since in \cite{bd96} a SPARC 1
machine was used which is slower by a factor $\approx 10$ than the 167 MHz
Sun ULTRA I used here, this means that our algorithms have comparable speeds.

\begin{table}
\begin{ruledtabular}
\caption{ Performances for the 3-d binary (HP-) sequences from
\protect\cite{yue95}.
} \label{table1}
\begin{tabular}{ccrrr}
 sequence &
 $-E_{\rm min}$\footnote{Ground state energies \protect\cite{yue95}.} &
 PERM\footnote{CPU times (minutes) per independent ground state hit,
      on 167 MHz Sun ULTRA I work station; from Ref.~\cite{bast98}} &
 nPERMss\footnote{CPU times, same machine} &
 nPERMis\footnote{CPU times, same machine} \\
   nr.    &          &         &        &        \\ \hline
    1     &    32    &    6.9  &   0.66 &   0.63 \\
    2     &    34    &   40.5  &   4.79 &   3.89 \\
    3     &    34    &  100.2  &   3.94 &   1.99 \\
    4     &    33    &  284.0  &  19.51 &  13.45 \\
    5     &    32    &   74.7  &   6.88 &   5.08 \\
    6     &    32    &   59.2  &   9.48 &   6.60 \\
    7     &    32    &  144.7  &   7.65 &   5.37 \\
    8     &    31    &   26.6  &   2.92 &   2.17 \\
    9     &    34    & 1420.0  & 378.64 &  41.41 \\
   10$\;\;$ &  33    &   18.3  &   0.89 &   0.47 \\
\end{tabular}
\end{ruledtabular}
\end{table}

{\bf (b)} Next we studied the two 2-d HP-sequences of length $N=100$ of
Ref.~\cite{rrp97}. They were originally thought to have ground states fitting
into a $10\times 10$ square with energies -44 and -46 \cite{rrp97}, but in
\cite{bast98} configurations fitting into this square were found with lower
energies, and moreover it was found that the configurations with lowest
energies ($E=-47$ resp.  $E=-49$) did not fit into this square. In the
present work we studied only configurations of the latter type.

For the second of these sequences, new lowest energy configurations with
$E=-50$
were found later in \cite{chikenji}, within 50 h CPU time on a 500 MHz DEC
21164A.
We now hit this energy 7 times, with an average CPU time of 5.8 h on a
667 MHz DEC 21264 between any two hits.

For the first sequence of Ref.~\cite{rrp97} we now hit several hundred
times states with $E=-48$, with ca.~2.6 min CPU time between successive hits.
One of these configurations is shown in Fig.~2.

\begin{figure}
  \begin{center}
   \psfig{file=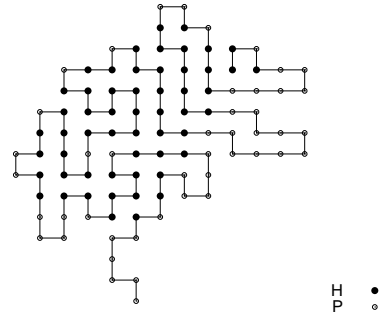,width=5.0cm,
angle=270}
   \vskip 0.3 true cm
\caption{Typical configuration with $E=-48$ of the first sequence of
Ref.~\cite{rrp97}.}
\label{2d-100}
\end{center}
\end{figure}

{\bf (c)} Several 2-d HP-sequences were introduced in \cite{um93}, where the
authors tried to fold them using a genetic algorithm. Except for the shortest
chains they were not successful, but putative ground states for all of them
were found in \cite{bast98,irbaeck,chikenji}. But for the longest of these
chains (with $N=64$, see Fig.~1), the ground state energy $E_{\rm min}=-42$ was
found in
\cite{bast98} only by means
of special tricks which amount to non-blind search. With blind search, the
lowest energy reached by PERM was -39. We should stress that PERM as used in
\cite{bast98} was blind for all cases except this 64-mer, in contrast to
wrong statements made in \cite{lw01}.

We now found putative ground states for all chains of \cite{um93} with blind
search. For the 64-mer the average CPU time per hit was ca. 30h on the
DEC 21264, which seems to be roughly comparable to the CPU times
needed in \cite{chikenji,irbaeck}, but considerably slower than \cite{bd96}.
As we already said in the introduction,
this sequence is particularly difficult for any growth
algorithm, and the fact that we now found it easily is particularly
noteworthy.

On the other hand, nPERMis was much faster than \cite{bd96} for the sequence
with $N=60$ of \cite{um93}. It needed $\approx 10$ seconds on the DEC 21264
to hit $E_{\rm min}=-36$, and $\approx 0.1$ second to hit $E=-35$. In contrast,
$E=-36$ was never hit in \cite{bd96}, while it took 97 minutes to hit $E=-35$.

{\bf (d)} A 85-mer 2-d HP sequence was given in \cite{konig}, where it was
claimed to have $E_{\rm min}=-52$. Using a genetic algorithm, the authors
could find only conformations with $E\ge -47$. In Ref.~\cite{lw01}, using
a newly developed {\it evolutionary Monte Carlo} (EMC) method, the authors
found the putative ground state when assuming large parts of its known
structure as constraints. This amounts of course to non-blind search. 
Without these constraints, the putative ground state was not hit in 
\cite{lw01} either, although the authors claimed their algorithm to be more 
efficient than all previous ones.

\begin{figure}
  \begin{center}
   \psfig{file=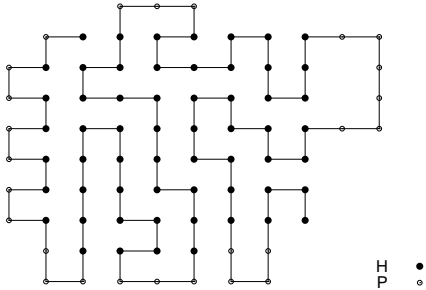,width=5.0cm,
angle=270}
   \vskip 0.3 true cm
\caption{New putative ground state configuration with $E=-53$ of the 2-d $N=85$
   chain taken from \cite{konig,lw01}.}
\label{2d-85}
\end{center}
\end{figure}

Both with nPERMss and with nPERMis we easily found states with $E=-52$, but
we also found many conformations with $E=-53$. For nPERMis at $\exp(1/T)=90$
it took ca. 10 min CPU time between successive hits on the Sun ULTRA 1.
One of those conformations is shown in Fig.~\ref{2d-85}.

{\bf (e)} As two easy cases we studied the two longest sequences from
\cite{kd01},
since we can compare there with CPU times given in \cite{kd01} for three
versions of a supposedly very efficient genetic algorithm. These 2-d HP
sequences with lengths $N=33$ and 48 have ground state energies -14 and -23,
respectively. In \cite{kd01}, the most efficient version needed on average
$\approx 45$min CPU (on an unspecified machine) to reach a ground state of 
the 33-mer.  For the 48-mer only energy -22 could be reached, within 
$\approx 2.5$h per hit. Using $\exp(1/T)=40$, it took the Sun ULTRA 1 just 
0.4 sec to hit one ground state of
the 33-mer, 7 sec to hit $E=-22$ for the 48-mer, and 16 min to hit a ground
state of the 48-mer. Thus the present algorithm is roughly 1000 times faster 
than that of \cite{kd01}.

\begin{figure}
 \begin{center}
   \psfig{file=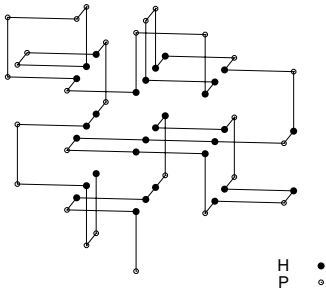,width=5.0cm,
angle=270}
\caption{Configuration with $E=-44$ of the $N=58$ HP sequence modeling protein
   BPTI from Ref.~\cite{dfc93,tt96}.}
\label{toma58}
\end{center}
\end{figure}

\begin{figure}
 \begin{center}
   \psfig{file=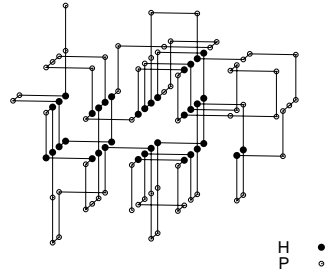,width=5.0cm,
angle=270}
\caption{Configuration with $E=-54$ of the $N=103$ HP sequence modeling
   cytochrome c from Ref.~\cite{lfd94,tt96}.}
\label{toma103}
\end{center}
\end{figure}

\begin{figure}
 \begin{center}
   \psfig{file=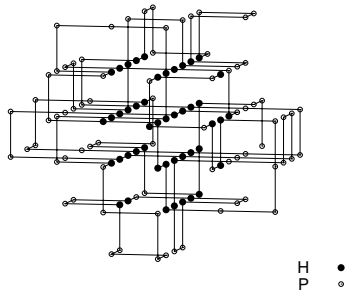,width=5.0cm,
angle=270}
\caption{Configuration with $E=-71$ of the $N=124$ HP sequence modeling
   ribonuclease A from Ref.~\cite{lfd94,tt96}.}
\label{toma124}
\end{center}
\end{figure}

\begin{figure}
 \begin{center}
   \psfig{file=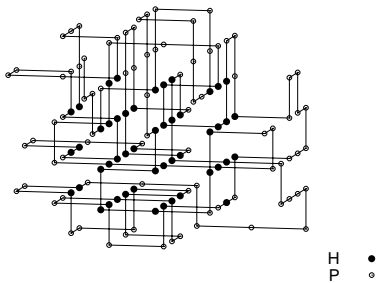,width=5.0cm,
angle=270}
\caption{Configuration with $E=-80$ of the $N=136$ HP sequence modeling
   a staphylococcal nuclease fragment, from Ref.~\cite{lfd94,tt96}.}
\label{toma136}
\end{center}
\end{figure}

{\bf (f)} Four 3D HP sequences with $N=58$, $103$, $124$, and $136$ were
proposed in \cite{dfc93,lfd94} as models for actual proteins or protein 
fragments. Low energy states for these sequences were searched 
in \cite{tt96} using a newly developed and supposedly very 
efficient algorithm. The energies reached in \cite{tt96} were $E=-42$, 
$-49$, $-58$ and $-65$, respectively. With nPERMis, we now found lower 
energy states after only few minutes CPU time, for all four chains. For
the longer ones, the true ground state energies are indeed {\it much} lower 
than those found in \cite{tt96}, see Table~\ref{toma}. Examples of the 
putative ground state configurations are shown in Figs.~\ref{toma58} to 
Fig.~\ref{toma136}.

Note the very low temperatures needed to fold the very longest chains in 
an optimal time. If we would be interested in excited states, higher 
temperatures would be better. For instance, to find $E=-66$ for the 
136-mer (which is one unit below the lowest energy reached in \cite{tt96}), 
it took just 2.7 seconds/hit on the DEC 21264 when using $\exp(1/T)=40$.

\begin{table}
\caption {\label{toma} Performance for the 3-d HP sequences from \cite{lfd94} }
\begin{ruledtabular}
\begin{tabular}{rrrrr}
  $N$  & $E_{\rm min}$\footnote{Lowest energies found in Ref.~\cite{tt96}}  &
  $E_{\rm min}$\footnote{Present work, using nPERMis} &
  $\exp(1/T)$ & CPU time
  \footnote{CPU times (hours) per independent lowest state hit,
  on 667 MHz DEC ALPHA 21264} \\
\hline
58  & -42   & -44  &  30  &  0.19  \\
103 & -49   & -54  &  60  &  3.12  \\
124 & -58   & -71  &  90  & 12.3$\;\; $  \\
136 & -65   & -80  & 120  & 110.$\;\;\;\; $  \\

\end{tabular}
\end{ruledtabular}
\end{table}

\begin{table*}
\caption {\label{table3} Newly found lowest energy states for binary
      sequences with interactions
$\vec{\epsilon}=(\epsilon_{HH},\epsilon_{HP},\epsilon_{PP})$.}
\begin{ruledtabular}
\begin{tabular}{rrccrc}
 &  &  & & old $E_{\rm min}$ &  \\
N & $d$ & $\vec{\epsilon}$ & Sequence & new $E_{\rm min}$ & Ref.\\
\hline
100 & 2 & -(1,0,0) &
$P_6HPH_2P_5H_3PH_5PH_2P_2(P_2H_2)_2PH_5PH_{10}PH_2PH_7P_{11}H_7P_2HPH_3P_6HPH_2$
& -47 &\cite{g98} \\
   &   &          &      & -48     &   \\
85 & 2 & -(1,0,0) &
$H_4P_4H_{12}P_6H_{12}P_3H_{12}P_3H_{12}P_3HP_2H_2P_2H_2P_2HPH$
& -52 & \cite{lw01} \\
   &   &          &      & -53  & \\
58 & 3 & -(1,0,0) &
$PHPH_3PH_3P_2H_2PHPH_2PH_3PHPHPH_2P_2H_3P_2HPHP_4HP_2HP_2H_2P_2HP_2H$
& -42 & \cite{tt96} \\
   &   &          &     & -44  & \\
103 & 3 & -(1,0,0) &
$P_2H_2P_5H_2P_2H_2PHP_2HP_7HP_3H_2PH_2P_6HP_2HPHP_2HP_5H_3P_4H_2PH_2P_5H_2P_4$
& -49 &
\cite{tt96} \\
   &   &          &  $H_4PHP_8H_5P_2HP_2$   &  -54 & \\
124 & 3 & -(1,0,0) &
$P_3H_3PHP_4HP_5H_2P_4H_2P_2H_2P_4HP_4HP_2HP_2H_2P_3H_2PHPH_3P_4H_3P_6H_2P_2HP_2$
& -58
& \cite{tt96} \\
  &    &         & $HPHP_2HP_7HP_2H_3P_4HP_3H_5P_4H_2PHPHPHPH$     &  -71 & \\
136 & 3 & -(1,0,0) &
$HP_5HP_4HPH_2PH_2P_4HPH_3P_4HPHPH_4P_{11}HP_2HP_3HPH_2P_3H_2P_2HP_2HPHPHP_8H$
& -65 &
\cite{tt96} \\
  &    &         &  $P_3H_6P_3H_2P_2H_3P_3H_2PH_5P_9HP_4HPHP_4$   &  -80 & \\

\end{tabular}
\end{ruledtabular}
\end{table*}

{\bf (g)} Several 3-d HP sequences were studied in \cite{yd95}, where also
their
{\it exact} ground state energies were calculated using the `constrained
hydrophobic core construction' (CHCC) which is essentially an exact enumeration
method tailored specifically to HP sequences on the cubic lattice. According to
\cite{yd95}, CHCC can be used to find all exact ground state configurations
for chains of length $N \approx 70$ to 88, depending on their degeneracies.

The longest chains given explicitly in \cite{yd95} together with their native
configurations are a four helix bundle with $N=64$ and $E_{\rm min} = -56$, and
a chain with $N=67$ folding into a configuration resembling an $\alpha/\beta$
barrel with $E_{\rm min} = -56$, too. Both have low degeneracy.

\begin{figure}
 \begin{center}
   \psfig{file=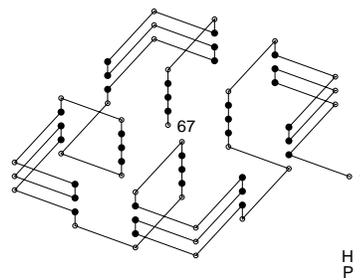,width=5.0cm,
angle=270}
\caption{Ground state configuration ($E=-56$) of the $N=67$ HP sequence given
in
   \cite{yd95}. It forms a structure resembling an $\alpha/\beta$ barrel. When
   starting at monomer \#67 ($\beta$ sheet end), nPERMis could find it easily,
   but not when starting from monomer \#1.}
\label{dill67}
\end{center}
\end{figure}

Finding ground states for the 64-mer was no problem for nPERMis. For
$\exp(1/T)=50$,
the DEC ALPHA 21264 machine needed on average 26.8 min CPU time to hit one of
them.
Things are a bit more interesting for the 67-mer. One of its ground states is
shown in Fig.~\ref{dill67}. Assume we want to let this grow, starting from the
$\beta$ sheet end (monomer \#67). Then we see that we always can form
immediately
stabilizing H-H bonds, and that we would be never seriously misled if we would
place
monomers greedily, at positions where they have low energies. Indeed, starting
from this end we had no problems with nPERMis: It took on average 67 min to hit
a native state on the DEC ALPHA 21264.

On the other hand, when starting with monomer \#1, we were unsuccessful and the
lowest energy reached was $E=-53$, even after much longer CPU times. This is
easily understood from Fig.~\ref{dill67}: starting from this end we have to go
repeatedly into directions which seem very unnatural at first sight, and which
get stabilized much later.

Notice that the difference between the two growth directions is not that there
is a folding nucleus when starting from \#67, and no folding nucleus when
starting from \#1. After the first quarter is built up, both give the {\it
same}
$\alpha/\beta$ pair. Building this first quarter is no problem even when
starting from \#1, at least when we use $C\ll 1$ (in which case it is built
essentially by complete enumeration). Thus the existence of a nucleus in the
traditional sense is not sufficient. Instead it is crucial that further growth
from this nucleus does not lead into false minima of the energy landscape.

{\bf (h)} Next we studied the two-species 80-mer with interactions (-1,0,-1)
that was introduced in \cite{tp92}. It was constructed in \cite{tp92} such as
to fold into a four helix bundle with $E=-95$, but two configurations with
$E=-98$ were found in \cite{g98} which essentially are $\beta$ sheet dominated.
These configurations were hit on average once every 80 hours on a 167 MHz Sun
ULTRA 1.
Later they were also found in \cite{chikenji2}, with similar CPU time as far as
we can tell. With nPERMis we needed only 5.3 hours / hit, on the same Sun ULTRA
1
(and for $8 \le \exp(1/T) \le 12$).

\begin{figure}
 \begin{center}
   \psfig{file=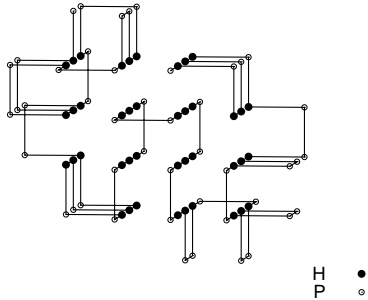,width=5.6cm,
angle=270}
\caption{Ground state configuration ($E=-72$) of the $N=88$ HP sequence given
in
   \cite{bd96}. It also forms a structure resembling an $\alpha/\beta$ barrel,
   with the core (the 4 $\beta$ strings) built from the central part of the
   chain. Without this core being already present, folding from neither end is
   easy.}
\label{beut88}
\end{center}
\end{figure}

{\bf (i)} Finally we also studied the 3-d HP sequence of length 88 given in
\cite{bd96}. As shown there, it folds into an irregular $\beta/\alpha$-barrel
with $E_{\rm min}=-72$. This is the only chain whose ground state we could {\it
not} find by our method, instead we only reached $E=-69$. This is in contrast
to the CG method which could find the lowest energy easily \cite{bd96}. The
difficulties of PERM with this sequence are easily understood by looking at
one of the ground states, see Fig.~\ref{beut88}. The nucleus of the hydrophobic
core is formed by amino acids \#36-53. Before its formation, a growth algorithm
starting at either end has to form very unstable and seemingly unnatural
structures which are stabilized only by this nucleus,
a situation similar to that in Fig.~1. In order to fold also
this chain, we would have either to start from the middle of the chain (as
done in \cite{bast98} for some sequences) or use some other heuristics which
help formation of the hydrophobic core. Since we wanted our algorithm to be
as general and ``blind" as possible, we did not incorporate such tricks.
The CG method, in contrast, is based on constructing an estimate of the
hydrophobic core and the hydrophilic shell,
and letting the chain grow to fill both in an optimal way,
using a heuristic cost function.

\medskip

Before leaving this section we should say that for all chains studied in this
paper we found also states
with $E=E_{\rm min}+1, E_{\rm min}+2,\ldots$. Thus none of the sequences showed
an energy gap above the (putative or exact) ground state. If such a gap is
indeed
typical for good folders, then none of the above sequences should be considered
as good folders.

A list containing all sequences for which we found new lowest energy
configurations
is given in Table~\ref{table3}.

\section{Discussion}

In the present paper we presented two new versions of PERM which is a
depth-first
implementation of the `go-with-the-winners' strategy (or sequential importance
sampling with re-sampling). The main improvement is that we now do not make
{\it identical clones} of high weight (partial) configurations, but we branch
such that each continuation is forced to be different. We do not expect this to
have much influence for systems at high temperatures, but as we showed, it
leads to
substantial improvement at very low temperatures. The two versions differ in
using
simple sampling (nPERMss) resp. importance sampling (nPERMis) when choosing
among
possible branches.

Although the method could be used for a much wider range of applications (see
\cite{chemnitz} for applications of PERM), we applied it here only to lattice
heteropolymers with two types of monomers. These represent toy models of
proteins,
and we hope that our results will also foster applications to more realistic
protein models. We showed only results for lowest energy configurations, but we
should stress that PERM and its new variants are not only optimization
algorithms.
They also give information on the full thermodynamic behaviour. We skipped this
here since finding ground states is the most difficult problem in general, and
sampling excited states is easy compared to it.

Comparing our results to previous work, we see that we found the known lowest
energy states in {\it all} cases but one. Moreover, whenever we could compare
with
previous CPU times, the comparison was favourable for our new algorithms,
except
for the CG method of Beutler and Dill \cite{bd96}. But we should stress that
the
latter is very specific to HP chains, uses strong heuristics regarding the
formation of a hydrophobic core, and does not give correct Boltzmann weights
for
excited states. All that is not true for our method. In general
nPERMis did slightly better than nPERMss, although the difference was much less
than a priori expected.

In principle, essentially the same algorithms can also be used for off-lattice
systems. This was already true for the original version of PERM which performed
well for Lennard-Jones polymers at temperatures around the $\Theta$-transition
\cite{hg95}, but rather badly for collapsed heteropolymers at
temperatures much below the $\Theta$ temperature \cite{fg-unpub}. Work is
presently in progress to see whether the new versions of PERM perform better,
and whether they can be used efficiently to study protein folding with
realistic interactions.

Acknowledgements:
We are indebted to Ralph Andrzejak for useful discussions and for critically
reading the manuscript. WN was supported by the DFG (SFB 237).

\section*{Appendix}

In this appendix we shall collect some basic facts about random sampling,
when {\it tuples} of instances are selected instead of individual instances.
The discussion will be very general. On the other hand, we will {\it not}
deal with problems specific to {\it sequential} sampling, i.e. we will
assume that we sample only for the choice of a single item (e.g. for the
placement of a single monomer).

Our central aim is thus to estimate a partition sum
\be
   Z = \sum_{i=1}^N q_i
\ee
where the {\it importances} $q_i$ might e.g. be Boltzmann-Gibbs factors,
and where $N$ is assumed to be finite (the generalization to infinite $N$
and to integrals instead of sums is straightforward). A conventional
Monte Carlo (MC) procedure consists in choosing `instances'
$i(\alpha),\;\alpha = 1,2,\ldots $ with probabilities $p_{i(\alpha)}$
such that each instance gives an unbiased estimate $\hat{Z}_1(\alpha)$ 
(the index ``1" will be explained in a minute). Thus, given $M$ such 
instances and letting $M$ tend to infinity, we have
\be
   Z = \lim_{M\to\infty} {1\over M} \sum_{\alpha=1}^M \hat{Z}_1(\alpha).
\ee
One easily sees that
\be
   \hat{Z}_1(\alpha) = {q_{i(\alpha)} \over p_{i(\alpha)}}       \label{hatZ1}
\ee
does the job. Indeed,
\be
   \lim_{M\to\infty} {1\over M} \sum_{\alpha=1}^M {q_{i(\alpha)} \over
p_{i(\alpha)}}
        = \sum_{i=1}^N p_i {q_i \over p_i} = Z .
\ee
At the same time we can also estimate the variance of $\hat{Z}_1$. We have
\begin{eqnarray*}
   {\rm Var} \hat{Z}_1 & = & \langle \hat{Z}_1^2 \rangle - \langle
\hat{Z}_1\rangle^2\\
       & = & \sum_{i=1}^N p_i \left({q_i \over p_i}\right)^2 - Z^2 \\
       & = & \sum_{i=1}^N {q_i^2\over p_i} - Z^2 .
\end{eqnarray*}
Up to now everything is correct for any choice of the probabilities $p_i$.
They get fixed e.g. by $p_i = 1/N$ (uniform sampling) or by demanding
${\rm Var} \hat{Z}_1$ to be minimal, under the constraint $\sum_i p_i =1$.
This simple variational problem gives $p_i^{\rm opt} \propto q_i$ which is
known
as {\it importance sampling}. For perfect importance sampling one
finds furthermore that ${\rm Var} \hat{Z}_1 = 0$.

Let us now assume that we select each time not one instance but $K$ instances,
all
of which are different. This requires of course $K\le N$. Moreover we will
assume
$K<N$, since otherwise this would amount to an exact summation of $Z$.
An advantage of such a strategy should be that we obtain a more widely and
uniformly spread sample. When $N\gg K$, this should not have a big effect, but in
our applications both $N$ and $K$ are small and the effect is substantial.

Thus each {\it event} consists in choosing a K-tuple $\{i_1,i_2,\ldots,i_K\}$,
with the $i_j$ mutually different,
from some probability distribution $p_{i_1,i_2,\ldots,i_K}$. We consider
tuples related by permutations as identical, i.e. without loss of generality
we can assume that $i_1<i_2<\ldots<i_K$. Each choice $\alpha$ of a tuple
$\{i_1(\alpha),i_2(\alpha), \ldots,i_K(\alpha)\}$ will lead to an estimate
$\hat{Z}_K(\alpha)$. Instead of Eq.(\ref{hatZ1}) we have now
\be
   \hat{Z}_K(\alpha) = {N \sum_{k=1}^K q_{i_k(\alpha)} \over K {N\choose K}
              p_{i_1(\alpha)\ldots i_K(\alpha)}}\;\;,
    \label{hatZK}
\ee
since one verifies easily that $\langle \hat{Z}_K(\alpha)\rangle = Z$.

The variance of $\hat{Z}_K(\alpha)$ is calculated just like that of
$\hat{Z}_1(\alpha)$,
\be
   {\rm Var} \hat{Z}_K  =  {N-1\choose K-1}^{-2}
        \sum_{i_1<\ldots <i_K}
        {(\sum_{k=1}^K q_{i_k})^2 \over p_{i_1\ldots i_K}} -Z^2\;.
\ee
Importance sampling is again obtained by minimizing it with respect to
$p_{i_1\ldots i_K}$, giving the result
\be
   p_{i_1\ldots i_K}^{\rm opt} = {N-1\choose K-1} {\sum_{k=1}^K q_{i_k} \over
              \sum_{j=1}^N q_j} \;.
\ee
The variance of $\hat{Z}_K$ vanishes again for this choice.

On the other hand, for uniform (or ``simple") sampling, with 
\be
   p_{i_1\ldots i_K}^{\rm ss} = {N \choose K}^{-1}\;,
\ee
we obtain 
\be
   {\rm Var} \hat{Z}_K = {(N-K) N^2 \over K(N-1)} {\rm Var} \,q \;.\qquad{\rm (simple
\; sampling)}
\ee
For $K=1$ this is the obvious result ${\rm Var} \hat{Z}_1 = N^2 {\rm Var}\, q$,
while for $K=N$ it gives ${\rm Var} \hat{Z}_N = 0$ as it should. For 
general $1<K<N$ the factor $1/K$ is trivial and results from the fact 
that each event corresponds to $K$ instances, while the factor $(N-K)/(N-1)$
gives the non-trivial improvement due to the fact that only {\it different}
instances are chosen in each event.

Finally, when using Eq.(\ref{hatZK}) for sequential sampling, one has to
attribute weights to each individual instance, instead of giving a weight
only to the entire tuple. The obvious solution is
\be
   W_{i_k(\alpha)} = { q_{i_k(\alpha)} N \over K {N\choose K}
              p_{i_1(\alpha)\ldots i_K(\alpha)}}\;.
\ee


\begin{thebibliography}{30}
\bibitem{footnote1} This prediction is biased (it is a lower bound to the true
   weight $W_n$, since it would be the weight if all Boltzmann factors were
   equal to 1). But this is not important for the following. In contrast, for
   nPERMis (see below), we will used an unbiased prediction, except for the
   replacement $q_\alpha \to \tilde{q}_\alpha$.
\end{thebibliography}
\end{document}